\documentclass[12pt]{article}
\usepackage{amssymb}
\begin{document}
\baselineskip=18pt
\pagenumbering{arabic}
\parskip1.5em
\newcommand{\ee}{{{\eta}\over {2}}}
\newcommand{\beq}{\begin{equation}}
\newcommand{\eeq}{\end{equation}}
\newcommand{\beqa}{\begin{eqnarray}}
\newcommand{\eeqa}{\end{eqnarray}}
\newcommand{\beqan}{\begin{eqnarray*}}
\newcommand{\eeqan}{\end{eqnarray*}}
\newcommand{\half}{{{1}\over{2}}}
\newcommand{\ihalf}{{{i}\over{2}}}
\newcommand{\quar}{{{1}\over{4}}}
\newcommand{\la}{\lambda}
\newcommand{\si}{\sigma}
\newcommand{\Si}{\Sigma}
\newcommand{\tf}{\Theta}
\newcommand{\3}{{\ss}}
\newcommand{\bra}{\langle 0|}
\newcommand{\ket}{|0\rangle}
\newcommand{\id}{{1}\hspace{-0.3em}{\rm{I}}}
\newcommand{\tn}[1]{T^{#1}}
\newcommand{\tr}{\bigtriangleup}
\newcommand{\trb}{\bar{\bigtriangleup}}
\def\b{\beta}
\def\d{\delta}
\def\g{\gamma}
\def\a{\alpha}
\def\s{\sigma}
\def\t{\tau}
\def\l{\lambda}
\def\La{\Lambda}
\def\e{\epsilon}
\def\r{\rho}
\def\d{\delta}
\def\wid{\widehat}
\def\ds{\displaystyle}
\def\be{\begin{equation}}
\def\ee{\end{equation}}
\def\beq{\begin{eqnarray}}
\def\eeq{\end{eqnarray}}
\def\ov{\overline}
\def\om{\omega}
\thispagestyle{empty}
\begin{flushright}BN--TH--2000--06\\
\end{flushright}
\vskip2.5em
\begin{center}
{\Large{\bf Polarization-free generators for the Belavin model}}
\vskip1.5em
T.-D. Albert\hspace{3em}K. Ruhlig\\\vskip3em
{\sl Physikalisches Institut der Universit\"at Bonn}\\
{\sl Nu{\ss}allee 12, D--53115 Bonn, Germany}\\\vskip1.5em
\end{center}
\vskip2em
\begin{abstract}
\noindent Employing a change of basis, the so-called factorizing Drinfel'd twist, we
construct polarization-free and completely symmetric creation operators
for a face type model equivalent to the Belavin model. A resolution of the
nested structure of the Bethevectors is achieved.
\end{abstract}
{\small{
e-mail: t-albert@th.physik.uni-bonn.de\\
\hspace*{1.3cm}ruhlig@th.physik.uni-bonn.de}}
\vfill\eject\setcounter{page}{1}

\section{Introduction}
The realm of integrable systems in 2d statistical physics or (1+1)d QFT has attracted much attention for a long time. Important contributions in this development were the solution of the anisotropic $sl(2)$ Heisenberg model ($XYZ$-model) by means of the Bethe ansatz \cite{baxter1} and its subsequent reformulation in an algebraic version, the algebraic Bethe ansatz or so called Quantum Inverse Scattering Mathod (QISM) \cite{fadtakh}. Generalizations to higher rank groups, which allow to treat models with internal degrees of freedom, were in this context achieved in \cite{kulresch}. The efforts resulted in a plethora of models soluble by means of the QISM, insofar that eigenvalues and eigenvectors of the Hamiltonian were found explicitely. This method also stimulated the investigation of various areas in mathematical physics, such as quantum groups, theory of knots  etc. Despite the indisputable achievements of the QISM and the rather simple action of the inverse problem operators, which can be interpreted as creation and annihilation operators for quasiparticles and as generating functions for the conserved quantities respectively, the study of correlation functions and formfactors has proven to be rather intricate. This is partly due to the fact that the solution of the inverse problem (expressing the original microscopic operators by means of the operators figuring in the algebraic Bethe ansatz) has only been achieved recently \cite{maillet, qis1, qis2}, and by the fact that the action of the quasiparticle  creation and annihilation operators which figure in the construction of the eigenfunctions (Bethe wavevectors) is obscured on the level of the microscopic variables (spin raising and lowering operators) by nonlocal effects arising from polarization clouds (compensating pairs of local raising and lowering spin operators). In a seminal  paper  Maillet  and  Sanchez de Santos  \cite{ms} revealed an application of so called Drinfel'd twists to obtain a basis for the $sl(2)$ $XXX$ and $XXZ$ model which allows to express the  creation and annihilation operators in a completely symmetric way with the further advantage of being polarization free, that is being built from the respective quasiclassical Gaudin operators dressed diagonally, thus supressing non-local effects for these operators. A generalized transformation has subsequently been applied to the $sl(2)$ $XYZ$ model \cite{abfpr} and been used to resolve the nested hierarchy in the Bethevectors of the $sl(n)$ $XXX$ model \cite{abfr}.\\
In this paper we report the construction of suitable Drinfel'd twists for the $\mathbb{Z}_n\times\mathbb{Z}_n$ symmetric Belavin model \cite{belavin}, which can be thought of as a n-state generalization of the $sl(2)$ $XYZ$ model. Our results generalize the findings for the $sl(n)$ $XXX$ model and provide a completely symmetric representation of the creation operators as well as a resolution of the hierarchical structure of the Bethe wavevectors.\par\noindent
The paper is organized as follows: Section 2 provides a short survey of the Belavin model and its reformulation as an face-type model. Section 3 deals with the factorizing twists and the computation of the operator valued entries of the monodromy matrix. In section 4 we discuss the Bethe wavevectors and section 5 contains the conclusions. 
\section{Belavin model and the corresponding IRF model}
A possible n-state generalization of the eight-vertex model \cite{baxter1} is given by the $\mathbb{Z}_n\times\mathbb{Z}_n$ symmetric model of Belavin \cite{belavin}, whose Boltzmann weights fulfill the Yang--Baxter equation (YBE) \cite{bovier}, \cite{cherednik}
\beqa
S(z_1-z_2)^{i_1,i_2}_{k_1,k_2}\,S(z_1-z_3)^{k_1,i_3}_{j_1,k_3}\,S(z_2-z_3)^{k_2,k_3}_{j_2,j_3}=S(z_2-z_3)^{i_2,i_3}_{k_2,k_3}\,S(z_1-z_3)^{i_1,k_3}_{k_1,j_3}\,S(z_1-z_2)^{k_1,k_2}_{j_1,j_2}\label{ybe1}
\eeqa
(here and subsequently double indices $k$ mean summation over $0,1,\ldots,n-1$ unless stated otherwise). It can be parametrized in the following way \cite{rt} \footnote{The slight difference compared to \cite{rt} originates from the normalization of the S-matrix.}
\beqa
S(z)_{i,j}^{k,l}=\left\{
\begin{array}{ll}
0\;\;\;\quad \quad \quad \quad \quad \quad \mbox{if}\; i+j\neq k+l\\
{{h(w)h(z)}\theta^{(i-j)}(z+w)\over{h(z+w)\theta^{(i-k)}(w)\theta^{(k-j)}(z)}}\;\quad \mbox{if}\; i+j=k+l;\mbox{mod}\,n \\
\end{array}
\right.\label{s}
\eeqa
with 
\beq
h(z)=\prod_{i=0}^{n-1}\theta^{(i)}(z)\left(\prod_{i=1}^{n-1}\theta^{(i)}(0)\right)^{-1}
\eeq
where $\theta^{(i)}(z)=\theta {\half-{i\over{n}} \choose\half}(z)$ represents the theta function of rational characteristics $1/2-i/n,1/2$. The theta function of characteristics $a,b\in\mathbb{R}$ is given as a Fourier series ($\tau$ is a fixed complex number in the upper half plane, and $\La_{\tau}\equiv\mathbb{Z}+\mathbb{Z}\tau$ is the lattice generated by $1$ and $\tau$)
\beq
\theta {a \choose b} (z)=\sum_{m\in\mathbb{Z}} exp\left(\pi \,i \tau(m+a)z+2\pi\,i(m+a)(z+b)\right)
\eeq
and has zeros at $z=(\half-b)+(\half-a)\tau\;\;\mbox{mod}\;\La_{\tau}$.\\
The matrix (\ref{s}) is unitary, $S(z)S(-z)=\id$ and obeys the initial condition $S(0)=P$ \cite{quano}, where $P$ is the permutation operator.\par\noindent
There exists a vertex-face map which transforms the Belavin model into a face type model \cite{jimbo1}, which in turn can be thought of as a multicomponent generalization of the six-vertex model.\\ 
The correspondence is given by 
\beq
S(z_1-z_2)M_a^{\mu}(z_1)\otimes M_{a+{\hat{\mu}}}^{\nu}(z_2)=M_{a+{\hat{\nu}}'}^{\mu'}(z_1)\otimes M_{a}^{\nu'}(z_2)R(z_1-z_2|a)^{\mu,\nu}_{\mu', \nu'}\label{vfm}
\eeq 
where $R(a|z)$ has the form \cite{zhou}
\beqa
R(z|a)^{\mu,\nu}_{\mu', \nu'}&=&b^{\mu, \nu}(z|a)\delta_{\mu'\mu}\delta_{\nu'\nu}+c^{\mu, \nu}(z|a)\delta_{\mu'\nu}\delta_{\nu'\mu}\label{R}\\
b^{\mu, \nu}(z|a)&=&{{h(z)}\over{h(z+w)}}{{h(s^{\nu}-s^{\mu}+w(a-{\hat{\nu}})^{\nu,\mu})}\over{h(s^{\nu}-s^{\mu}+w(a-{\hat{\nu}})^{\nu,\mu}+w)}}\prod_{i=0}^{n-1}{{g_{i\nu}(a)}\over{g_{i\nu}(a+{\hat{\mu}})}}\prod_{i=0}^{n-1}{{g_{i\mu}(a+{\hat{\nu}})}\over{g_{i\mu}(a)}}\nonumber\\
c^{\mu, \nu}(z|a)&=&{{h(z+w)}\over{h(z)}}{{h(s^{\nu}-s^{\mu}+w(a-{\hat{\nu}})^{\nu,\mu}+w+z)}\over{h(s^{\nu}-s^{\mu}+w(a-{\hat{\nu}})^{\nu,\mu}+w)}}\label{rmatrix}\\\nonumber\\
g_{i\mu}(a)&=&\left\{
\begin{array}{ll}
1\;\;\;\quad \mbox{if}\; i\geq\mu\\
h(s^0-s^{\mu}+wa^{0,\mu})\quad \mbox{if}\; i=0\\
h(s^i-s^{\mu}+wa^{i,\mu}-w)\quad \mbox{if}\; 0<i<\mu\\
\end{array}
\right.
\eeqa
where $s^{\mu}$ are arbitrary complex numbers, $a\in\sum_{\mu=0}^{n-1}\mathbb{C}\Lambda_{\mu}$ with $\Lambda_{\mu}$ weights of the affine Lie algebra $A_{n-1}^{(1)}$, ${\hat{\mu}}=\Lambda_{\mu+1}-\Lambda_{\mu},\;\mu=0,1,\ldots,n-1$ and $\Lambda_{0}=\Lambda_{n}$. The $a^{\mu,\nu}$ is given in \cite{jimbo1} and obeys $(a-\hat{\nu})^{\mu,\mu}=0$.\\ 
The intertwining vector $M^{\mu}_{a}(z)$ is given by
\beqa
M^{\mu}_{a}(z)&=&{}^t\left(\phi^{\mu}_{a}(z)_0,\ldots,\phi^{\mu}_{a}(z)_{n-1}\right)\prod_{i=0}^{n-1}g^{-1}_{i\mu}(a)\label{M}\\
\phi^{\mu}_{a}(z)_j&=&\theta^{(j)}(z-nwa^{\mu,0}).
\eeqa
Applying a sum of threefold tensorproducts of functions (\ref{M}) to (\ref{ybe1}) using the vertex face map (\ref{vfm}) and exploiting the explicit index structure of (\ref{R}) we obtain a modified Yang-Baxter equation 
\beqa
&&R(|z_1-z_2|a+{\hat{\a}}_3)^{\a_1,\a_2}_{\mu_1, \mu_2} R(z_1-z_3|a)^{\mu_1,\a_3}_{\b_1, \mu_3}R(z_2-z_3|a+{\hat{\b}}_1)^{\mu_2,\mu_3}_{\b_2, \b_3}\nonumber\\
&=&R(z_2-z_3|a)^{\a_2,\a_3}_{\mu_2, \mu_3}R(z_1-z_3|a+{\hat{\mu}}_2)^{\a_1,\mu_3}_{\mu_1, \b_3}R(z_1-z_2|a)^{\mu_1,\mu_2}_{\b_1, \b_2}\,.\label{ybe}
\eeqa
This can be written symbolically as (we set $R(z_i-z_j|a)=R_{ij}(a)$)
\beqan
R_{12}(a+{\hat{\sigma}}_3)R_{13}(a)R_{23}(a+{\hat{\sigma}}_1)=R_{23}(a)R_{13}(a+{\hat{\sigma}}_2)R_{12}(a)\,.
\eeqan
The monodromy matrix for a chain with $N$ sites associated with the R-matrix (\ref{R}) is
\beq
T_{0,1\ldots N}(\l|a)=R_{0N}(a+{\hat{\sigma}}_1+\ldots+{\hat{\sigma}}_{N-1})\ldots R_{02}(a+{\hat{\sigma}}_1)R_{01}(a)\label{monod}
\eeq
where $0$ denotes the horizontal auxiliary space with spectral parameter $\l$; $1,\ldots,N$ label the vertical quantum spaces (with local inhomogeneities $\left\{z_i\right\}$) whose tensorproduct constitutes the physical Hilbertspace ${\cal{H}}_N$.  An algebraic construction of eigenvalues of the monodromy matrix associated to the matrix (\ref{s}) was performed in \cite{zhou} and parallels the procedure for the eight-vertex model in \cite{fadtakh} (One has to take into account that our construction of the monodromy matrix (\ref{monod}) differs from the monodromy matrix in \cite{zhou} by an additional transformation in the physical space).\\
Subsequently we will focus on the construction of a factorizing $F$ matrix for the R-matrix (\ref{R}) and pursue its consequences for the structure of the monodromy matrix (\ref{monod}).
\section{The F basis}
The factorizing $F$-matrix for $N$ sites ($N$ quantum spaces), being defined by the relation $R^{\sigma}_{1\ldots N}(a)=F_{\sigma(1\ldots N)}^{-1}(a)F_{1\ldots N}(a)$ \cite{ms}, turns out to be given by formally the same expression as found in \cite{abfr} for the $sl(n)$ $XXX$-model
\beqa
F_{1\ldots N}(a)&=& \sum_{\sigma\in S_N}\sum_{\alpha}^{\quad \quad *}
\prod_{i=1}^N P_{\sigma(i)}^{\alpha_{\sigma(i)}}R_{1\ldots N}^{\sigma_{{\alpha}}}(z_{1},\ldots,z_{N}|a)\label{Fmatrix}
\eeqa
where $P_{i}^{\alpha_i}$ projects on the ${\alpha_i}$-th component in the $i$-th space and the labels $\a_{\sigma(i)}$ satisfy the conditions 
\beqa
\alpha_{\sigma(i+1)}&\geq&\alpha_{\sigma(i)}\quad \mbox{if}\quad\sigma(i+1)>\sigma(i)\nonumber\\
\alpha_{\sigma(i+1)}&>&\alpha_{\sigma(i)}\quad \mbox{if}\quad\sigma(i+1)<\sigma(i)\;.\label{cond}
\eeqa
The modification of the Yang-Baxter equation (\ref{ybe}) requires a particular rule for the handling of the parameter $a$ in the formation of the intertwining matrix $R^{\sigma}(a)$
(related to the permutation $\sigma$), which can be read off from the modified composition law
\beqa
R^{\sigma\sigma_{i}}(a)&=&R_{\sigma(i),\sigma(i+1)}(\tilde{a}_i)R^{\sigma}(a)\nonumber\\
\tilde{a}_i&=&a+{\hat{\sigma}}_{\sigma(1)}+\ldots +{\hat{\sigma}}_{\sigma(i-1)}\label{complaw}
\eeqa
where $\sigma_{i}$ is the transposition of $i,i+1$, and $\sigma$ an arbitrary permutation.\\
$R^{\sigma}(a)$ has the intertwining property 
\beqa
R^{\sigma}(a)T_{0,1\ldots N}(a)=T_{0,\sigma(1)\ldots \sigma(N)}(a)R^{\sigma}(a+{\hat{\sigma}}_h)\ \label{perm}
\eeqa
where $\s_h$ is associated with the matrix indices in the space 0.\\
The matrix $F_{1\ldots N}(a)$ satisfies the factorizing equation 
\beqa
R^{\sigma}_{1\ldots N}(a)=F_{\sigma(1\ldots N)}^{-1}(a)F_{1\ldots N}(a)\;.\label{ffe}
\eeqa
A proof of the latter equation can be found in \cite{abfr} and relies on the fact that 
\beq
P_{i}^{\alpha}P_{j}^{\alpha}R_{ij}=P_{i}^{\alpha}P_{j}^{\alpha}\id_{ij}
\;.\label{ppr}
\eeq
The modification of the composition law induced by the parameter $a$ being insignificant.\\
We will proceed by computing elements of the monodromy matrix in the new basis provided by the F-matrix. The transformation law ${\tilde{T}}_{0,1\ldots N}(a)=F_{1\ldots N}(a)T_{0,1\ldots N}(a)F_{1\ldots N}^{-1}(a+{\hat{\sigma}}_h)$ is enforced by the requirement that the resulting operator is symmetric, i.e. $\tilde T_{0,1\ldots N}(a)=\tilde T_{0,\sigma(1)\ldots \sigma(N)}(a)$, which then follows from (\ref{perm}).\\
\\
The computation of the diagonal element $T_{n-1\,n-1}(\l|a)$ proceeds along the same lines as that in the generalized $sl(n)$ $XXX$-model in \cite{abfr}. Let us consider the action of the matrix $F(a)$ on $T_{n-1\,n-1}(a)$:
\beqa
F_{1\ldots N}(a)T_{n-1\,n-1}(a)&=&\sum_{\s\in S_N}\sum^{\quad\quad *}_{\a_{\s(1)}\ldots\a_{\s(N)}}
\prod_{i=1}^N P_{\s(i)}^{\a_{\s(i)}} R_{1\ldots N}^{\s}(a)P^{n-1}_0 T_{0,1\ldots N}(a) P^{n-1}_0\nonumber\\
&=&\sum_{\s\in S_N}\sum^{\quad\quad *}_{\a_{\s(1)}\ldots\a_{\s(N)}}
\prod_{i=1}^N P_{\s(i)}^{\a_{\s(i)}}P^{n-1}_0 T_{0,\s(1)\ldots \s(N)}(a) P^{n-1}_0 R_{1\ldots N}^{\s}(a+\s_h)\,.\label{Tnn1}
\eeqa
The specialization to the entry $(n-1,n-1)$ of the auxiliary space here is achieved by the projectors $P^{n-1}_0$. For the second equality in (\ref{Tnn1}) we have used relation (\ref{perm}) and the obvious fact that $P^{n-1}_0$ commutes with $R^{\s}_{1\ldots N}(a)$. To simplify the following argument we distinguish in the sum $\sum^*$ cases of various multiplicities of the occurrence of the group index $n-1$: 
\beqa
F_{1\ldots N}(a)T_{n-1\,n-1}(a)&=&\sum_{\s\in S_N}\sum_{k=0}^N\sum^{\quad\quad *'}_{\a_{\s(1)}\ldots\a_{\s(N)}}\prod_{j=N-k+1}^N P_{\s(j)}^{n-1}\delta_{\a_{\s(j)},n-1}\prod_{j=1}^{N-k} P_{\s(j)}^{\a_{\s(j)}}\nonumber\\
&&\times P^{n-1}_0 T_{0,\s(1)\ldots \s(N)}(a) P^{n-1}_0
 R_{1\ldots N}^{\s}(a+\s_h)\,.\label{Tnn2}
\eeqa
Let us consider the prefactor of $R^{\s}_{1\ldots N}(a+\s_h)$ on the r.h.s. of Eq. (\ref{Tnn2}) more closely. Using specific features of the $R$-matrices we can rewrite it as follows:
\beqa
&&\prod_{j=1}^{N-k} P_{\s(j)}^{\a_{\s(j)}}\prod_{j=N-k+1}^N P^{n-1}_{\s(j)} P^{n-1}_0\,T_{0,\s(1)\ldots \s(N)}(a)\,P^{n-1}_0\nonumber\\
&=&\prod_{j=1}^{N-k} P_{\s(j)}^{\a_{\s(j)}}\left(R_{0,\s(N)}(a+\sum_{m=1}^{N-1}\hat{\s}_{\s(m)})\right)_{n-1n-1}^{n-1n-1} \ldots\left(R_{0,\s(N-k+1)}(a)\right)_{n-1n-1}^{n-1n-1}\nonumber\\
&&\times P^{n-1}_0\,T_{0,\s(1)\ldots \s(N-k)}(a)\,P^{n-1}_0\prod_{j=N-k+1}^N P^{n-1}_{\s(j)}\nonumber\\
&=&\prod_{j=1}^{N-k} P_{\s(j)}^{\a_{\s(j)}}P^{n-1}_0\,T_{0,\s(1)\ldots \s(N-k)}(a)\,P^{n-1}_0 \prod_{j=N-k+1}^N P^{n-1}_{\s(j)}\nonumber\\
&=&\prod_{i=1}^{N-k}\left(R_{0\s(i)}(a+\sum_{m=1}^{i-1}\hat{\s}_{\s(m)})\right)_{n-1,\a_{\s(i)}}^{n-1,\a_{\s(i)}}\prod_{j=1}^{N-k} P_{\s(j)}^{\a_{\s(j)}}\prod_{j=N-k+1}^N P^{n-1}_{\s(j)}\,P^{n-1}_0\label{prp}
\eeqa
Inserting the r.h.s. of (\ref{prp}) into Eq. (\ref{Tnn2}) one sees that the product $\prod_i\left(R_{0\s(i)}(a+\sum_{m=1}^{i-1}\hat{\s}_{\s(m)})\right)_{n-1,\a_{\s(i)}}^{n-1,\a_{\s(i)}}$ creates a diagonal dressing factor for $\tilde{T}_{n-1\,n-1}(a)$ and the product of projectors applied to $R^{\s}(a+\s_h)$ gives $F_{1\ldots N}(a+\s_h)$. One obtains
\beq
\tilde T_{n-1\,n-1}(\l|a) =Y_{n-1}(a) \otimes_{i=1}^N \mbox{diag}\{b(\l-z_i),\ldots,b(\l-z_i),1\}\,.         \label{T1}
\eeq
with the abbreviations
\beqa
Y_j(a)&=&\prod_{m=0}^{j-1}\left\{\prod_{i=0}^{n-1}{{g_{ij}(\widetilde{a_m}+k_m\hat{m})}\over{{g_{ij}(\widetilde{a_m})}}}\prod_{j_m=1}^{k_m}\left[\prod_{i=0}^{n-1}{{g_{im}(\widetilde{a_m}+(j_m-1)\hat{m})}\over{{g_{im}(\widetilde{a_m}+(j_m-1)\hat{m}+\hat{j})}}} \right.\right.\times\nonumber\\
&& \left.\left.\times{{h(s^m-s^j+w[\widetilde{a_m}+(j_m-2)\hat{m}]^{m,j})}\over{h(s^m-s^j+w[\widetilde{a_m}+(j_m-2)\hat{m}]^{m,j}+w)}}\right]\right\}\nonumber\\
\widetilde{a_m}&=&a+\sum_{i=0}^{m-1}k_i\,\hat{i}\nonumber\\
b(\l)&=&{{h(\l)}\over{h(\l+w)}}
\eeqa
and $k_m$ gives the multiplicity of the upper matrix labels $\a_i=m$.\\
\\
To compute $T_{n-1\,n-2}(\l|a)$ one has to distinguish in the sum $\sum^*$ cases of various multiplicities $k_{n-1}$ and $k_{n-2}$ of the occurrence of group indices $n-1$ and $n-2$:
\beqa
F_{1\ldots N}(a)T_{n-1\,n-2}(a)&=&\sum_{\s\in S_N}\sum_{k_{n-1}=0}^N\sum_{k_{n-2}=0}^{N-k_{n-1}}\sum^{\quad\quad *''}_{\a_{\s(1)}\ldots\a_{\s(N)}}\prod_{j_{n-1}=N-k_{n-1}+1}^N P_{\s(j_{n-1})}^{n-1}\prod_{j_{n-2}=N-k_{n-1}-k_{n-2}+1}^{N-k_{n-1}} P_{\s(j_{n-2})}^{n-2}\nonumber\\
&&\times\prod_{j=1}^{N-k_{n-1}-k_{n-2}} P_{\s(j)}^{\a_{\s(j)}} P^{n-1}_0 T_{0,\s(1)\ldots \s(N)}(a) P^{n-2}_0
 R_{1\ldots N}^{\s}(a+\s_h)\,.\label{C11}
\eeqa
Evaluating the matrix product in $T_{0,\s(1)\ldots \s(N)}(a)$ leads to
\beqa
&&\prod_{j_{n-1}=N-k_{n-1}+1}^N P_{\s(j_{n-1})}^{n-1}\prod_{j_{n-2}=N-k_{n-1}-k_{n-2}+1}^{N-k_{n-1}} P_{\s(j_{n-2})}^{n-2}\prod_{j=1}^{N-k_{n-1}-k_{n-2}} P_{\s(j)}^{\a_{\s(j)}} P^{n-1}_0 T_{0,\s(1)\ldots \s(N)}(a) P^{n-2}_0\nonumber\\
&=&\sum_{i=N-k_{n-1}-k_{n-2}+1}^{N-k_{n-1}}\left(R_{0\widetilde{N-k_{n-1}}}(\hat{a}_{N-k_{n-1}})\right)^{n-1\,n-2}_{n-1\,n-2}\ldots\left(R_{0\widetilde{i+1}}(\hat{a}_{i+1})\right)^{n-1\,n-2}_{n-1\,n-2}\nonumber\\
&&\times\left(R_{0\tilde{i}}(\hat{a}_{i})\right)^{n-1\,n-2}_{n-2\,n-1}\left(R_{0\widetilde{i-1}}(\hat{a}_{i-1})\right)^{n-2\,n-2}_{n-2\,n-2}\ldots\left(R_{0\widetilde{N-k_{n-1}-k_{n-2}+1}}(\hat{a}_{N-k_{n-1}-k_{n-2}+1})\right)^{n-2\,n-2}_{n-2\,n-2}\nonumber\\
&&\times\prod_{k=1}^{N-k_{n-1}-k_{n-2}}\left(R_{0\tilde{k}}(\hat{a}_{k})\right)^{n-2\,\a_{\tilde{k}}}_{n-1\,\a_{\tilde{k}}}E^{\tilde{i}}_{n-2,n-1}\prod_{j\neq i}P_{\tilde{j}}^{\a_{\tilde{j}}}P_{\tilde{i}}^{n-1}E^{0}_{n-1,n-2}\nonumber\\
&=& \sum_{i=N-k_{n-1}-k_{n-2}+1}^{N-k_{n-1}}\prod_{l=i+1}^{N-k_{n-1}}b^{n-1,n-2}(\l-z_{\tilde{l}}|\hat{a}_l)c^{n-2,n-1}(\l-z_{\tilde{i}}|\hat{a}_i)\nonumber\\
&&\times\prod_{k=1}^{N-k_{n-1}-k_{n-2}}b^{n-2,\a_{\tilde{k}}}(\l-z_{\tilde{k}}|\hat{a}_k)E^{\tilde{i}}_{n-2,n-1}\prod_{j\neq i}P_{\tilde{j}}^{\a_{\tilde{j}}}P_{\tilde{i}}^{n-1}E^{0}_{n-1,n-2}\label{C12}
\eeqa
with the abbreviations
\beqa
\hat{a}_i&=&a+\sum_{m=1}^{i-1}\hat\s_{\tilde{m}}\nonumber\\
\tilde{i}&=&\s(i)\nonumber\\
\left(E^i_{a,b}\right)^{\a_i}_{\b_i}&=&\d_{\a_i,a}\d_{\b_i,b}\, .\nonumber
\eeqa
\\
One notes that in the calculation the index $\a_{\tilde{i}}$ has changed from $n-2$ to $n-1$. As the distribution of $\a$'s is therefore no longer consistent with the conditions (\ref{cond}) in the sum $\sum^*$ one has to correct this fact by commuting the site $\tilde{i}$ through all higher sites $\tilde{j}$ with $\a_{\tilde{j}}=n-2$. So, taking into account (\ref{complaw}), one has to insert an additional factor
\beq
R_{\tilde{i}\,\widetilde{N-k_{n-1}}}(\hat{a}_{\widetilde{N-k_{n-1}}}+\s_h)\ldots R_{\tilde{i}\,\widetilde{i+1}}(\hat{a}_{\widetilde{i+1}}+\s_h)\label{corc1}
\eeq
between the projectors and $R_{1\ldots N}^{\s}(a+\s_h)$ in (\ref{C11}). Because of Eq. (\ref{ppr}) no further corrections are neccessary. For the following calculation two equations are needed
\beq
P^{n-1}_i\,P^{n-2}_j\,\id_{ij}=P^{n-1}_i\,P^{n-2}_j\left\{b^{n-1,n-2}(z_i-z_j|a)^{-1}R_{i\,j}(a)-{{c^{n-2,n-1}(z_i-z_j|a)}\over{b^{n-1,n-2}(z_i-z_j|a)}}P_{i\,j}\right\}\label{C1R}
\eeq
and
\beq
E^i_{n-1,n}\,P^{n-1}_j\,P_{i\,j}=E^j_{n-1,n}\,P^{n-1}_i\label{C1E}
\eeq
with $\left(P_{i\,j}\right)^{\a_i\,\a_j}_{\b_i\,\b_j}=\d_{\a_i,\b_j}\d_{\a_j,\b_i}$ the permutation operator in space i and j.\\
\\
Let us now concentrate on the term with $i=N-k_{n-1}-k_{n-2}+1$ in (\ref{C12}) and use (\ref{C1R}) to create the needed factor (\ref{corc1}). Because of (\ref{C1E}) the second term in (\ref{C1R}) gives rise to an $E^{\tilde{j}}_{n-2,n-1}$ with $\tilde{j}\neq \tilde{i}$. So the only possibility to get an  $E^{\tilde{i}}_{n-2,n-1}$ is to use the first term in (\ref{C1R}). Corrections in the other terms with $j>N-k_{n-1}-k_{n-2}+1$ cannot lead to an expression with $E^{\tilde{i}}_{n-2,n-1}$ as $\tilde{j}$ has not to be commuted with the site $\tilde{i}$. So the only term that contains $E^{\tilde{i}}_{n-2,n-1}$ after the corrections for that special $R^{\s}_{1\ldots N}$ in (\ref{C12}) is
\beqa
&&\prod_{l=i+1}^{N-k_{n-1}}{{b^{n-1,n-2}(\l-z_{\tilde{l}}|\hat{a}_l)}\over{b^{n-1,n-2}(z_{\tilde{i}}-z_{\tilde{l}}|\hat{a}_l)}}c^{n-2,n-1}(\l-z_{\tilde{i}}|\hat{a}_i)\prod_{k=1}^{N-k_{n-1}-k_{n-2}}b^{n-2,\a_{\tilde{k}}}(\l-z_{\tilde{k}}|\hat{a}_k)E^{\tilde{i}}_{n-2,n-1}\nonumber\\
&=&\prod_{l=i+1}^{N-k_{n-1}}{{b(\l-z_{\tilde{l}})}\over{b(z_{\tilde{i}}-z_{\tilde{l}})}}c^{n-2,n-1}(\l-z_{\tilde{i}}|\tilde{a}_{n-2})\prod_{k=1}^{N-k_{n-1}-k_{n-2}}b^{n-2,\a_{\tilde{k}}}(\l-z_{\tilde{k}}|\hat{a}_k)E^{\tilde{i}}_{n-2,n-1}.\label{C1sim}
\eeqa
Because of the symmetry of $\tilde{T}_{0,1\ldots N}(\l|a)$ all other terms have to be of the same form as (\ref{C1sim}). After taking into account the projectors the resulting expression is
\\
\beqa
\tilde{T}_{n-1\,n-2}(\l|a)&=&Y_{n-2}(a)\sum_{i=1}^Nc^{n-2,n-1}(\l-z_i|\tilde{a}_{n-2})E^i_{n-2,n-1}\nonumber\\
&& \otimes^N_{j\neq i}\,\mbox{diag}\{b(\l-z_j),\ldots,b(\l-z_j),b(\l-z_j)b^{-1}(z_i-z_j),1\}.\label{T21a}
\eeqa
\\
For the calculation of $\tilde{T}_{n-1\,n-3}(\l|a)$ one has to distinguish the cases $n-1$, $n-2$ and $n-3$ in the sum $\sum^*$. The only difference compared to $\tilde{T}_{n-1\,n-2}(\l|a)$ is a term containing a product $E^i_{n-3,n-2}\,E^j_{n-2,n-1}$ now showing up in the matrix product in $T_{0,\s(1)\ldots \s{N}}(a)$. One again has to correct the distribution of $\a's$ with the analog of the equations (\ref{C1R}) and (\ref{C1E})  
\beq
P^{\a_i}_i\,P^{\a_j}_j\,\id_{ij}=P^{\a_i}_i\,P^{\a_j}_j\left\{b^{\a_i,\a_j}(z_i-z_j|a)^{-1}R_{i\,j}(a)-{{c^{\a_j,\a_i}(z_i-z_j|a)}\over{b^{\a_i,\a_j}(z_i-z_j|a)}}P_{i\,j}\right\}\label{C2R}
\eeq
\beq
E^i_{a,b}\,P^a_j\,P_{i\,j}=E^j_{a,b}\,P^{a}_i\label{C2E}
\eeq
and also with a new relation which has to be taken into account when dealing with the term containing the product $E^i_{n-3,n-2}\,E^j_{n-2,n-1}$ :
\beq
E^i_{n-3,n-1}\,P^{n-2}_j\,P_{i\,j}=E^i_{n-3,n-2}E^j_{n-2,n-1}.\label{C2P}
\eeq
This reasoning leads to ($b_{i\,k}=b(z_i-z_k)$):
\beqa
\tilde{T}_{n-1\,n-3}(\l|a)&=&Y_{n-3}(a)\sum_{i=1}^Nc^{n-3,n-1}(\l-z_i|\tilde{a}_{n-3})E^i_{n-3,n-1}\nonumber\\
&& \otimes^N_{k\neq i}\mbox{diag}\{b(\l-z_k),\ldots,b(\l-z_k),b(\l-z_k)b^{-1}_{i\,k},b(\l-z_k)b^{-1}_{i\,k},1\}\nonumber\\
&& +Y_{n-3}(a)\sum_{i\neq j}^Nf_2(n-3,n-2;n-1)E^i_{n-3,n-2}\,E^j_{n-2,n-1}\nonumber\\
&& \otimes^N_{k\neq i,j}\mbox{diag}\{b(\l-z_k),\ldots,b(\l-z_k),b(\l-z_k)b^{-1}_{i\,k},b(\l-z_k)b^{-1}_{j\,k},1\}
\eeqa
with
\beqa
f_2(n-3,n-2;n-1)&=&c^{n-2,n-1}(\l-z_j|\tilde{a}_{n-2})c^{n-3,n-2}(\l-z_i|\tilde{a}_{n-3})\nonumber\\
&-&{{c^{n-2,n-1}(z_i-z_j|\tilde{a}_{n-2})}\over{b^{n-1,n-2}(z_i-z_j|\tilde{a}_{n-2})}}b^{n-1,n-2}(\l-z_j|\tilde{a}_{n-2})c^{n-3,n-1}(\l-z_i|\tilde{a}_{n-3})\nonumber\\
\label{f2}
\eeqa
where the second term in (\ref{f2}) has its origin in the second term in (\ref{C2R}) and in ({\ref{C2P}).\par\noindent
Proceeding in an analogous manner one obtains in the general case
\beqa
\tilde T_{n-1\;\a}(\l|a)&=&\sum_{k=1}^{\a}\;\sum_{i_1\ne\ldots\ne i_k}
\;\;\sum_{n-1-\a<n_1<\ldots n_k<n-1}\nonumber\\\nonumber\\
&&E^{(i_1)}_{n_1,n_2}\otimes E^{(i_2)}_{n_2,n_3}\otimes\ldots\otimes E^{(i_k)}_{n_k,n-1}\;\;f_k(n_1,n_2,\ldots,n_k;n-1)\;Y_{n-1-\a}(a)\nonumber\\\nonumber\\
&&\hspace{-5em}\otimes_{j\neq\{i_k\}}\;\mbox{diag}\left\{b(\l-z_j),\ldots,{\underbrace{b(\l-z_j)b^{-1}_{i_1 j},\ldots,b(\l-z_j)b^{-1}_{i_1 j}}_{n_2-n_1}},\ldots,{\underbrace{b(\l-z_j)b^{-1}_{i_k j},\ldots,b(\l-z_j)b^{-1}_{i_k j}}_{n-1-n_k}},1 \right\}\nonumber\\
\label{Ta}
\eeqa
with $f_k(n_1,n_2,\ldots,n_k;n_0)$ being defined recursively by \footnote{In the rational $sl(n)$ case this recursion relation leads directly to the result in Eq. (42) in \cite{abfr}.}
\beqa
f_k(n_1,n_2,\ldots,n_k;n_0)=-{{c^{n_k,n_0}_{i_{k-1},i_k}}\over{b^{n_0,n_k}_{i_{k-1},i_k}}} b^{n_0,n_k}_{0,i_k}f_{k-1}(n_1,n_2,\ldots,n_{k-1};n_0) +c^{n_k,n_0}_{0,i_k}f_{k-1}(n_1,n_2,\ldots,n_{k-1};n_k)
\eeqa
where
\beqan
f_1(n_1;n_0)=c^{n_1,n_0}_{0,i_{n_{1}}}
\eeqan
{\underline{Remark}}: As the above procedure to obtain the operators of the monodromy matrix in the F-basis did not rely on the invariance of the monodromy matrix under combined $sl(n)$ rotations in the auxilliary and quantum space, the $sl(n)$  XXZ-model can be treated in the same way, and we obtain expressions for the $\tilde T_{n-1\;\a}$ similar to that in (\ref{Ta}), with the diffenence that the $a$ dependence vanishes, and the parametrization of the elements of the R-matrix is trigonometric instead of elliptic  ($c(\l)={{sinh(w)}\over{sinh(\l+w)}},\;b(\l)={{sinh(\l)}\over{sinh(\l+w)}}$).

\section{Bethe wavevectors}
Our presentation of the hierarchical Bethe ansatz will be rather sketchy. For details we refer the reader to \cite{kulresch, zhou}. \\
The operators $T_{n-1\,\a}$ $(\a<n-1)$ act as quasiparticle creation operators  and satisfy the Faddeev--Zamolodchikov algebra
\beqa
T_{n-1\,\a}(\l|a)T_{n-1\,\b}(\mu|a+\hat{\a})=T_{n-1\,\gamma}(\mu|a)T_{n-1\,\delta}(\l|a+\hat{\g})R(\l-\mu|a)_{\a\,\b}^{\delta\,\gamma}\label{fzaR}
\eeqa
or in components of the R-matrix
\beqa
\left[T_{n-1\,\a}(\l|a),\,T_{n-1\,\a}(\mu|a+\hat{\a})\right]&=&0\label{fza1}\\
T_{n-1\,\a}(\l|a)T_{n-1\,\b}(\mu|a+\hat{\a})&=&{1\over{b^{\b\a}(\mu-\la|a)}}T_{n-1\,\b}(\mu|a)T_{n-1\,\a}(\l|a+\hat{\b})\nonumber\\
&-&{{c^{\b\a}(\mu-\la|a)}\over{b^{\b\a}(\mu-\la|a)}}T_{n-1\,\b}(\l|a)T_{n-1\,\a}(\mu|a+\hat{\b})\label{fza2}\,.
\eeqa
Inspired by \cite{kulresch}, an ansatz for a Bethe vector $\Psi_n$ is given in terms of a linear superposition of products of operators $T_{n-1\,\a}$ acting on a reference state $\Omega ^{(n)}_N$:
\beq
\Psi_n (N;\l_1,\ldots,\l_p|a)=\sum_{\a_1,\ldots,\a_{p}}
\Phi_{\a_1,\ldots,\a_{p}}
T_{n-1\,\a_1}(\l_1|a)T_{n-1\,\a_2}(\l_2|a+\hat{\a_1})\ldots T_{n-1\,\a_{p}}(\l_{p}|a+\sum_{i=1}^{n-1}\hat{\a_i})\,\Omega^{(n)}_N\nonumber\\
\label{Psi}
\eeq
where the  reference state $\Omega^{(n)}_N$ is constituted as a $N$-fold tensor product of lowest weight states $v_n^{(i)}={}^t(0,\ldots,0,1)$ in ${\mathbb{C}}_n^{(i)}$ 
\beqan
\Omega_N=\otimes_{i=1}^N v_n^{(i)}
\eeqan
and the $\Phi_{\a_1,\ldots,\a_{p}}$ denote some c-number coefficients, which themselves have to be chosen s.t. they are components of a $sl(n-1)$ wavevector, leading to a nested structure finally giving a $sl(2)$ eigenvalue problem.\\
It is important to note that the reference state is invariant under the $F$-transformation
\beqan
F\,\Omega^{(n)}_N=\Omega^{(n)}_N
\eeqan
due to the special form of the R-matrix.\\
We will not impose the Bethe ansatz equations for the spectral parameters $\{\l_i\}$ which turns the vector (\ref{Psi}) into an eigenvector of the transfer matrix, that is we consider the Bethe wavevector being ``off-shell'' \cite{bab}.\\ 
In what follows we want to determine the functional form of such vectors, using the explicit form of the operators relevant for the Bethe wavevectors.\\
In order to clarify the arguments employed in the course of the computation we will present the case of $sl(2)$ and $sl(3)$ in quite a detail. The generalization to the general case of $sl(n)$, $n>3$ will then be rather straightforward.\\  
For the $sl(2)$ case we have from (\ref{T21a})
\beqa
\tilde{T}_{10}(\l|a)&=&\sum_{i=1}^N c^{0,1}(\l-z_i|a)\s_+^{(i)}
\otimes_{j\ne i}
{\left(\matrix{
b(\l-z_j)b_{ij}^{-1}&0\cr
0\quad\quad&1\cr
}\right)}_{[j]}\label{T10}\\
\tilde{\Psi}_2(N;\l_1,\ldots,\l_p|a)&=&\tilde{T}_{10}(\l_1|a)\tilde{T}_{10}(\l_2|a+\hat{0})\ldots \tilde{T}_{10}(\l_p|a+(p-1)\cdot\hat{0})\,\Omega^{(2)}_N\nonumber\\\nonumber\\
&=&\sum_{i_1\ne\ldots\ne i_{p}} B_{p}^{(2)}(\l_1,\ldots,\l_{p}|z_{i_1},\ldots,z_{i_{p}}|a)
\s_+^{(i_1)}\ldots\s_+^{(i_{p})}\,\Omega^{(2)}_N\;.
\label{Psi_2}
\eeqa
The c-number coefficient $B^{(2)}(\{\l_i\}|\{z_i\}|a)$ is, due to the ``diagonally dressed'' spin raising operators $\s_{+}^{i}$ in (\ref{T10}), of the form 
\beqa
B^{(2)}_p(\l_1,\ldots,\l_p|z_{1},\ldots,z_{p})&=&
\sum_{\s\in S_p}\prod_{m=1}^p c^{0,1}(\l_m-z_{\s(m)}|a+(m-1)\cdot\hat{0})
\prod_{l=m+1}^{p}{{b(\l_{m}-z_{\s(l)})}\over{b(z_{\s(m)}-z_{\s(l)})}}.\nonumber\\
\label{B2}
\eeqa
We now turn to the $sl(3)$ case. The strategy will rely on the symmetry of the wavevector under the exchange of arbitrary spectral parameters (the verification of this fact follows the same lines as the one in \cite{takh} using (\ref{ybe}) and (\ref{fzaR})) which enables us to concentrate on a particularily simple term in the sum (\ref{Psi}), and the repeated use of the Faddeev--Zamolodchikov algebra. \\
These ideas lead us to propose the following form for the Bethe wavevector
\beqa
&&{\tilde\Psi}_3(N,\l_1,\ldots,\l_{p_0};\l_{p_0+1},\ldots,\l_{p_0+p_1}|a) \nonumber\\
&=&\sum_{\s \in S_{p_0}}B^{(2)}_{p_1}(\l_{p_0+1},\ldots,\l_{p_0 +p_1}|
\l_{\s(1)},\ldots,\l_{\s(p_1)}|a)\nonumber\\
&\times&\prod_{k=1}^{p_1}\prod_{l=p_1+1}^{p_0}
b^{1,0}(\l_{\s(k)}-\l_{\s(l)}|a+(k-1)\cdot\hat{0}+(l-p_1-1)\cdot\hat{1})^{-1}\nonumber\\
&\times&{\tilde T}_{21}(\l_{\s(p_1 +1)}|a)\ldots{\tilde T}_{21}(\l_{\s(p_0)}|a+(p_0-p_1-1)\cdot\hat{1})\nonumber\\
&\times&{\tilde T}_{20}(\l_{\s(1)}|a+(p_0-p_1)\cdot\hat{1})\ldots {\tilde T}_{20}(\l_{\s(p_1)}|a+(p_0-p_1)\cdot\hat{1}+(p_1-1)\cdot\hat{0})\,\Omega^{(3)}_N\, .
\label{Psi3a}
\eeqa
Consider a special term in the sum (\ref{Psi}) of the form (which is motivated by the fact that the associated coefficient $\Phi$ is especially simple to compute, see below)
\beqa
&&{\tilde T}_{20}(\l_1|a){\tilde T}_{20}(\l_1|a+\hat{0})\ldots {\tilde T}_{20}(\l_{p_1}|a+(p_1-1)\cdot\hat{0})\nonumber\\
&\times&{\tilde T}_{21}(\l_{p_1+1}|a+p_1\cdot\hat{0}){\tilde T}_{21}(\l_{p_1+2}|a+p_1\cdot\hat{0}+\hat{1})\ldots {\tilde T}_{21}(\l_{p_0}|a+p_1\cdot\hat{0}+(p_0-p_1-1)\cdot\hat{1})\,\Omega_N^{(3)}\, .\nonumber\\
\label{01}
\eeqa
Commuting all ${\tilde T}_{20}(\l_1|a)$ to the right using the first term in (\ref{fza2}) yields an additional factor
\beqan
\prod_{x=1}^{p_1}\prod_{y=p_1+1}^{p_0}\left\{b^{1,0}(\l_{y}-\l_{x}|a+(x-1)\cdot\hat{0}+(y-p_1-1)\cdot\hat{1})\right\}^{-1}\label{factorb}\,.
\eeqan
It has to be noted that the associated c-number coefficient  $\Phi^{(2)}_{0\ldots 01\ldots 1}$ in (\ref{01}) is not evaluated in the $sl(3)$ F basis. It can however be expressed in the form (\ref{B2}) as it is invariant under the action of the $sl(2)$ F-matrix. This is due to the fact that it constitutes a component of the $sl(2)$ vector whose labels (a non-decreasing series of $\a_i$ with respect to the original ordering of sites $i$) correspond via (\ref{cond}) to the identity permutation in the definition of the F-matrix (\ref{Fmatrix}).\\
Invoking the exchange symmetry we arrive thus at the formula (\ref{Psi3a}).\\
The creation operators with respect to the lowest weight state are of the form (cf. (\ref{Ta}))
\beqa
{\tilde T}_{21}(\l|a)\;&=&\;Y_1(a)\sum_{i=1}^N c^{1,2}(\l-z_i|\widetilde{a_1}) E_{1,2}^{(i)}\otimes_{j\neq i}
\mbox{diag}\{
b(\l-z_j),b(\l-z_j)b^{-1}_{ij},1\}_{[j]}\label{T21}\\
{\tilde T}_{20}(\l|a)\;&=&
\;\sum_{i=1}^N c^{0,2}(\l-z_i|\widetilde{a_0}) E_{0,2}^{(i)}\otimes_{j\neq i}
\mbox{diag}\{
b(\l-z_j)b^{-1}_{ij},b(\l-z_j)b^{-1}_{ij},1\}_{[j]}\nonumber\\
&+&\mbox{terms involving}\,E^{(i)}_{1,2}\otimes E^{(j)}_{0,1}\label{T20}\,.
\eeqa
The second term in (\ref{T20}) annihilates the vacuum $\Omega_N^{(3)}$, which is why we did not cite the explicit form of the prefactors accompanying these roots. The form of the creation operators permits us to further simplify the wavevector
(\ref{Psi3a}). Taking into account the action of the roots on the respective dressing as well as the fact that both groups of creation operators generate a factor similar to the $sl(2)$ problem, we obtain
\beqa
&&{\tilde\Psi}_3(N,\l_1,\ldots,\l_{p_0};\l_{p_0+1},\ldots,\l_{p_0+p_1}|a)
\nonumber\\
&=&\sum_{\s \in S_{p_0}}
\sum_{i_1\ne\ldots\ne i_{p_0}} 
\prod_{k=1}^{p_1}\prod_{l=p_1+1}^{p_0}b^{1,0}(\l_{\s(l)}-\l_{\s(k)}|a+(k-1)\cdot\hat{0}+(l-p_1-1)\cdot\hat{1})^{-1}\,\,b(\l_{\s(l)}-z_{i_k})\nonumber\\
&\times&B_{p_0-p_1}^{1,2}(\l_{\s(p_1+1)},\ldots,\l_{\s(p_0)}|z_{i_{p_1+1}},\ldots,z_{i_{p_0}}|a)\;
B_{p_1}^{0,1}(\l_{p_0+1},\ldots,\l_{p_0+p_1}|\l_{\s(1)},\ldots,\l_{\s(p_1)}|a)\nonumber\\\nonumber\\ 
&\times&B_{p_1}^{0,2}(\l_{\s(1)},\ldots,\l_{\s(p_1)}|z_{i_{1}},\ldots,z_{i_{p_1}}|a+(p_0-p_1)\cdot\hat{1})\;E_{1,2}^{(i_{p_1+1})}\ldots E_{1,2}^{(i_{p_0})} E_{0,2}^{(i_{1})}\ldots E_{0,2}^{(i_{p_1})} \Omega^{(3)}_N\nonumber\\
\label{Psi3b}
\eeqa
where we defined as a generalization of (\ref{B2})
\beqa
&&B^{\a,\b}_p(\l_1,\ldots,\l_p|z_{1},\ldots,z_{p}|a)\nonumber\\
&=&\sum_{\s\in S_p}\prod_{m=1}^p \left\{c^{\a,\b}(\l_m-z_{\s(m)}|\widetilde{a_{\a}}+(m-1)\cdot\hat{\a})\,Y_{\a}(a+(m-1)\cdot\hat{\a})\right\}
\prod_{l=m+1}^{p}{{b(\l_{m}-z_{\s(l)})}\over{b(z_{\s(m)}-z_{\s(l)})}}.\nonumber\\
\eeqa
We are now in the position to proceed to the general $sl(n)$ case.\\
We start with a special term of (\ref{Psi}), where all operators $\tilde{T}_{n-1\,\alpha}$ are to the left of operators $\tilde{T}_{n-1\,\beta}$ if $\alpha<\beta$. The associated coefficient $\Phi$ again remains invariant under the action of F, which enables us to express it in a manner similar to (\ref{B2}). We then use the Faddeev-Zamolodchikov algebra to reverse the order of all operators. Once again only the term containing one $E^i_{a,b}$ in the expression (\ref{Ta}) of the respective operators $\tilde{T}_{n-1\,\a}$ contributes in this special ordering.\\
The wavevector can be expressed in a form similar to (\ref{Psi3b}) ($\bar{\a}\equiv n-2-\a$)
\beqa
\tilde{\Psi}_n(N,p_0,p_1,\ldots,p_{n-2})=
\sum_{i_1\ne\ldots\ne i_{p_0}} B^{(n)}_{p_0,p_1,\ldots,p_{n-2}}
(\l_1,\ldots,\l_{p_0+...p_{n-2}}|z_{i_1},\ldots,z_{i_{p_0}}|a)
\prod_{\a=0}^{n-2}\prod_{j=p_{\bar{\a}+1}+1}^{p_{\bar{\a}}}
E_{\a\;n-1}^{(i_j)} \Omega^{(n)}_N\nonumber\\
\label{Psina}
\eeqa
with the recursion relation for $B^{(n)}$
\beqa
&&B^{(n)}_{p_0\,p_1\,\ldots\,p_{n-2}}(\l_1,\ldots,\l_{p_0+p_1+\ldots p_{n-2}}|z_{1},\ldots,z_{{p_0}}|a)\nonumber\\
&=&\sum_{\s\in S_{p_0}}\prod_{\a=0}^{n-3}\,\prod_{\b=\a+1}^{n-2}\quad\prod_{k_{\a}=p_{\bar{\a}+1}+1}^{p_{\bar{\a}}}
\quad\prod_{l_{\b}=p_{\bar{\b}+1}+1}^{p_{\bar{\b}}}b(\l_{\s(l_{\b})}-z_{{k_{\a}}})\nonumber\\\nonumber\\
&\times&b^{\b,\a}\left(\l_{\s(l_{\b})}-\l_{\s(k_{\a})}|a+\sum_{m=0}^{\a-1}(p_{\bar{m}}-p_{\bar{m}+1})\cdot\hat{m}+(k_{\a}-p_{\bar{\a}+1}-1)\cdot\hat{\a}+(l_{\b}-p_{\bar{\b}+1}-1)\cdot\hat{\b}\right)^{-1}\nonumber\\\nonumber\\
&\times&\prod_{\g=0}^{n-2} 
B^{\g,n-1}_{p_{\bar{\g}}-p_{\bar{\g}+1}}(\l_{\s(p_{\bar{\g}+1}+1)}\ldots\l_{\s(p_{\bar{\g}})}|
z_{{p_{\bar{\g}+1}+1}}\ldots z_{{p_{\bar{\g}}}}|a+\sum_{m=\g+1}^{n-2}(p_{\bar{m}}-p_{\bar{m}+1})\cdot\hat{m})\nonumber\\\nonumber\\
&\times&B^{(n-1)}_{p_1\ldots p_{n-2}}(\l_{p_0+1}\ldots \l_{p_0+p_1+\ldots +p_{n-2}}|
\l_{\s(1)}\ldots\l_{\s(p_1)}|a)
\label{Bnrec}
\eeqa
which can be solved explicitely to yield
\beqa
&& B^{(n)}_{p_0\,p_1\,\ldots\,p_{n-2}}(\l_1,\ldots,\l_{p_0+p_1+\ldots +p_{n-2}}|z_{1},\ldots,z_{{p_0}}|a)\nonumber\\
&&=\sum_{\s_0\in S_{p_0}}\sum_{\s_1\in S_{p_1}}\ldots\sum_{\s_{n-3}\in S_{p_{n-3}}}\prod_{i=0}^{n-3}\;\prod_{\a_i=0}^{n-3-i}\;\prod_{\b_i=\a_i+1}^{n-2-i}\;\prod_{k_{\a_i}=p_{\bar{\a_i}+1}+1}^{p_{\bar{\a_i}}}\;\prod_{l_{\b_i}=p_{\bar{\b_i}+1}+1}^{p_{\bar{\b_i}}}b(\l_{q_{i-1}+{\s_i(l_{\b_i})}}-\l_{q_{i-2}+\s_{i-1}(k_{\a_i})})\nonumber\\\nonumber\\
&\hspace{-3em}\times&\hspace{-2em}b^{\b_i,\a_i}\left(\l_{q_{i-1}+{\s_i(l_{\b_i})}}-\l_{q_{i-1}+{\s_{i}(k_{\a_i})}}|a+\sum_{m=0}^{\a_i-1}(p_{\bar{m}}-p_{\bar{m}+1})\cdot\hat{m}+(k_{\a_i}-p_{\bar{\a_i}+1}-1)\cdot\hat{\a_i}+(l_{\b_i}-p_{\bar{\b_i}+1}-1)\cdot\hat{\b_i}\right)^{-1}\nonumber\\\nonumber\\
&\hspace{-3em}\times&\hspace{-2.2em}\prod_{\gamma_i=0}^{n-2-i}\,B^{\g_i,n-1}_{p_{\bar{\g_i}}-p_{\bar{\g_i}+1}}(\l_{q_{i-1}+\s_i(p_{\bar{\g_i}+1}+1)}\ldots\l_{q_{i-1}+\s_i(p_{\bar{\g_i}})}|\l_{q_{i-2}+\s_{i-1}(p_{\bar{\g_i}+1}+1)}\ldots\l_{q_{i-2}+\s_{i-1}(p_{\bar{\g_i}})}|a+\sum_{m=\g_i+1}^{n-2-i}(p_{\bar{m}}-p_{\bar{m}+1})\cdot\hat{m})\nonumber\\\nonumber\\
&\hspace{-3em}\times&\hspace{-1.2em}B^{0,1}_{p_{n-2}}\left(\l_{q_{n-3}+1}\ldots\l_{q_{n-3}+p_{n-2}}|\l_{q_{n-4}+\s_{n-3}(1)}\ldots\l_{q_{n-4}+\s_{n-3}(p_{n-2})}|a\right)\nonumber\\\label{Bn}
\eeqa
where we defined
$$
q_{i}=\sum_{j=0}^{i}p_{j};\;\;q_{-1}=0
$$
and 
$$
\l_{\s_{-1}(k)}=z_k\,.
$$
By expressing the $sl(n)$ wavevector (\ref{Psina}) with the help of $sl(2)$ building blocks (\ref{Bn}) we have achieved a resolution of the hierarchy.\par\noindent
{\underline{Remark}}: Once again the arguments in this section apply to the $sl(n)$ $XXZ$-model as well, leading to a wavevector for this model which displays the same features as (\ref{Psina}) in connection with (\ref{Bn}), without the $a$ dependence and with a trigonometric parametrization.
\section{Conclusion}
We accomplished the construction of a factorizing F-matrix for the Belavin model enabling one to construct completely symmetric creation operators which moreover are devoid of non-local effects from polarization clouds. These operators were used to resolve the intricacies of the nested structure of Bethe wavevectors.\\
In contrast to the $sl(n)$ $XXX$-model the above method does not rely on an $sl(n)$ invariance of the monodromy matrix, which renders the extraction of the elements in the grid of the monodromy matrix much more cumbersome. The only ingredients needed in our computation are the form of the R-matrix (\ref{R}), i.e. its structure, unitarity and the fact that it constitutes a representation of the permutation group, and the property (\ref{ppr}). Thus our findings directly yield the corresponding expression for the generalized $sl(n)$ $XXZ$-model. \\
In view of the similarities between the generalized $sl(n)$ rational, trigonometric and elliptic model it is tempting to ask whether such a distinguished basis exists for all integrable models in two dimensions. In \cite{schroer} it was shown that in every integrable two-dimensional quantum field theory there exist semi-local polarization-free generators which are localized in wedge-shaped regions of Minkowski space. It is conceivable that there is a relation between these operators and polarization-free operators in lattice spin models.\\
\\
We hope that these results might prove useful for the construction of formfactors starting from the microscopical level.
\vspace{1.5cm}\\
{\bf{Acknowledgement:}} 
We thank H. Boos, R. Flume and R.H. Poghossian for helpful discussions.

\vfill\eject
\end{document}